\documentclass{svproc}
\usepackage{graphicx}
\usepackage{url}

\begin{document}
\mainmatter              
\title{Diffusion without Spreading of a Wave Packet in Nonlinear Random Models}
\titlerunning{Diffusion without spreading}  
%
\author{Serge Aubry
\authorrunning{S. Aubry} 
%
%
\institute{Laboratoire L\'eon Brillouin, CEA Saclay, 91191-Gif-sur-Yvette, France,\\
\email{serge.aubry91@gmail.com}}}

\maketitle              

\begin{abstract}
We discuss the long time behaviour of a finite energy wave packet in nonlinear Hamiltonians on infinite lattices at arbitrary dimension, exhibiting linear Anderson localization. Strong  arguments both mathematical and numerical, suggest for infinite models that small amplitude wave packets may generate stationary quasiperiodic solutions (KAM tori)  almost  undistinguishable from linear wave packets. The probability of this event  is  non vanishing at small enough amplitude and goes to unity at amplitude zero. Most other wave packets (non KAM tori) are chaotic. We discuss the Arnold diffusion conjecture (recently partially proven) and propose a modified Boltzmann statistics for wave packets valid in generic models. The consequence is that  the probability that a chaotic wave packet  spreads to zero amplitude is zero. It must always remain focused around one or few chaotic spots which moves randomly over the whole system and generates subdiffusion.We study a class of Ding Dong models also generating subdiffusion where the nonlinearities are replaced by hard core potentials. Then we prove rigorously that  spreading is impossible for any initial wave packet. 
\end{abstract}
\section{Introduction}
\label{sec:1}

It is well-known that wave propagation becomes impossible in linear random system with strong  enough disorder
because of Anderson localization \cite{And58}. Such a situation may also occur in other non random models which are for example incommensurate \cite{AA81}. Our purpose is to study  the same problem when nonlinearities are taken into account in theses models with a  purely discrete linear spectrum that is where all the linear eigen modes are square summable and  spatially localized. 

It has been suggested mostly on the base of numerical simulations and rough arguments  that Anderson localization is destroyed. A fully chaotic dynamics  is supposed to take place as a consequence of the extra nonlinearities which couple the localized linear modes one with  each other \cite{PS08,FKS09,SKKF09,LBKSF10}. Nevertheless, we also proved that large enough amplitude wave packet cannot spread in some cases like models with norm conservation \cite{KKFA08}. Otherwise, we suggested that small enough amplitude  wave packets  \cite{JKA10,Aub11} should  generate stationary quasiperiodic solutions (KAM tori) despite the number of degrees of freedom is infinite and moreover that the probability this happens goes to $1$ as the amplitude of the wave packet goes to zero. The initial wave packets which do  not generate  KAM tori, are chaotic as well as in finite systems. In infinite system, we argued that their spatial spreading should stop after some transient time \cite{Aub11} when wave packet get a small enough amplitude so that it reaches the region where  KAM tori are very dense.
However we could  not predict  the behaviour of the wave packet beyond this initial regime. The aim of this paper is to complete this work.  
But note first now that stopping spreading does not mean stopping diffusion as we explain now.

\subsection{Difference  between  Spreading and Diffusion}  Indeed, ``Spreading" and ``Diffusion" were generally considered by physicists as equivalent terms when considering the behaviour of waves packets. We apologize that we also did this confusion in our early works on the topics before we realized this  was not only a semantic problem. In order to remain correct with this updated definition, the word diffusion  should be replaced by spreading everywhere in all our early papers concerning this topics.

We say that a wave packet with amplitude $A(\mathbf{r},t)$ is spreading when its maximum amplitude $\sup_{\mathbf{r}}  |A(\mathbf{r},t)| $  goes to zero as time goes to infinity. Thus non spreading
means that the wave packet amplitude does not go uniformly to zero as $t\rightarrow +\infty$. This property  does not implies it has a non vanishing limit or even it has a limit. Actually, in  our situation, the wave packet has no limit.

Diffusion originally concerns particles moving randomly. At each time, the particle is localized at a random position $\mathbf{r}$ but we can define its probability density $\mathcal{P}(\mathbf{r},t)$ to be at position $\mathbf{r}$ over all possible realizations of the random walk  which is a   function of time $t$.  We say we have diffusion when $\sup_{\mathbf{r}} \mathcal{P}(\mathbf{r},t)$ goes to zero as time goes to infinity. 

Our problem here concerns the behaviour of initially localized wave packets  in nonlinear lattices at any finite dimension $d$  with sites labeled $\mathbf{i}\in \mathcal{Z}^d$ and discrete  linear spectrum.  We conclude  that in all cases,  the maximum value of its energy density $E_{\mathbf{i}} (t)$ does not go to zero at infinite time and thus  wavepackets are  never  spreading. However, for a given initial wave packet, the probability density  $\mathcal{P}(E_{\mathbf{i}} (t)) $ of its energy distribution with respect to all disorder realizations,  may go uniformly to zero as time goes to infinity, and so that we can say  we have wave packet diffusion without wave packet spreading. The purpose of this talk is to  explicit our arguments which hold in all lattice models where complete linear Anderson localization occurs.  We first need to go back to the foundation of statistical mechanics:

\subsection{Discussion of the Boltzmann Ergodic  Hypothesis}

It is assumed  that most Hamiltonian dynamical systems with many degrees of freedom  generates chaotic trajectories so that their physical behaviour cannot be described in a deterministic way.  The  well-known Ergodic Boltzmann Hypothesis (BEH) assumes that most trajectories generated by a Hamiltonian system, are chaotic and uniformly dense in the phase subspace at constant energy  (microcanonical ensemble) considered with the standard Liouville measure conserved by the Hamiltonian flow. The Boltzmann entropy  is then $k_B \ln W$  where $W$ is the accessible volume in the phase space. In other words, considering an arbitrary trajectory after a  long  time, in most cases the probability that it is found an arbitrary region of the phase space its simply proportional to its volume.  \textit{Most} means that the trajectories which do not fulfill  this property have zero Liouville measure inside this subspace and thus can be neglected in the statistics. Actually there are many special trajectories (e.g. periodic  or else ) among chaotic trajectories which are not ergodic but these trajectories are proven to have zero measure and can be neglected in the global statistics. This hypothesis is fundamental for constructing the well known theory of statistical mechanics.

This hypothesis  is by far not obvious. Indeed there are also special Hamiltonian systems (called integrable)  the trajectories of which are not chaotic but quasiperiodic and consequently are not ergodic systems. For example, commonly used harmonic systems are integrable.  However, it was believed  in such cases that  infinitely small hamiltonian perturbations completely breaks integrability and restore full ergodicity. Since the Hamiltonian of real systems cannot be perfectly known,  it is expected that generally arbitrarily small unknown perturbations restore the validity of statistical mechanics.  In any case,  a weak coupling with a thermal bath  (with a continuous spectrum)  forces thermalization.
Thus whatever is the initial state of a real physical system even far from thermal equilibrium, it is believed it must relax to a thermal equilibrium at some temperature after some transient time. In other words, the system maximizes its entropy.

Fermi even proposed a rigorous proof for ergodicity thus proving the validity of BEH for it but it was erroneous \cite{Poe01}.  Later, the Fermi Pasta Ulam Tsingou model raised an apparent paradox in chaos theory which is still under debate.  On contrary, the pioneering work of Kolmogorov \cite{Kol54} proved that the ergodic hypothesis was wrong at least for some finite Hamiltonian  near an integrable limit. Later,  Arnold and Moser extended  his results  and proved  that very generally integrability is not completely destroyed under small but arbitrary perturbations and consequently that  a finite Hamiltonian system  assumed  to be perfectly isolated may not  thermalize spontaneously.

Our problem of relaxation of a wave packet looks similar. We consider an \textit{infinite} system in its ground state in which we inject a \textit{finite}  energy as a focused wave packet. Consequently, if the system relaxes to  thermal equilibrium and since the energy density is zero, its temperature at  complete thermalization should be strictly zero. Thus the wave packet  should spread to zero which is the general belief. Actually,  we claim that the wave packet cannot spread completely but nevertheless  may become  diffusive according to the above definition.

The reason is that the  KAM theory initially  proven only for finite size system, may remain valid  in the vicinity of integrable Hamiltonian for infinite systems on lattices under two conditions which are: 
1 - The linearized spectrum is purely discrete,
2 - The energy of the wave packet is finite.

\section{Dynamics of Hamiltonians with a finite number of degrees of freedom}

First we have to recall as a brief review,  the  main results and  conjectures about KAM theory for perturbed integrable  systems which highlight the role of resonances (KAM is the acronym of the names of the main pioneering contributors: Kolmogorov, Arnold and Moser).  
For details, there are many  textbooks and reviews in the literature about  this topics.
 
 \subsection{KAM tori in finite nearly  integrable Hamiltonian systems}
Such systems exhibit quasiperiodic trajectory as well as chaotic trajectories.
 According to Liouville, an  integrable Hamiltonian involving  $n+n$ degrees of freedom (i.e. described by $n$ pairs of conjugate variables) is characterized by the existence of  $n$ independent time invariants. 
The Liouville-Arnold theorem states that under few  extra assumptions (like that the boundedness of the constant energy submanifolds  that is no trajectories go to infinity),  there exists a canonical change of variables which define a new set of conjugate variables $\{I_i,\theta_i\}$ where $I_i$ are real numbers called actions and  $\theta_i$ are angles defined modulo $2\pi$. With these new  variables, the Hamiltonian becomes  only a function $H_0(\{I_i\}$ of the actions independent of its conjugate variables.  Then, the Hamilton equations $\dot{\theta}_i= \partial H_0 /\partial I_i$,
$\dot{I}_i= -\partial H /\partial \theta_i =0$ imply that the actions $I_i$ are time invariant and the angles
are rotating uniformly with frequencies $\omega_i(\{I_j\})= \partial H_0 /\partial I_i$.  
Most trajectories of such a system are quasiperiodic on $n$ invariant dimensional tori with $n$ fundamental frequencies which  generally depend on the actions. 

However there is a dense subset of invariant tori called resonant when there exists a  set  $\{k_i\}$ of  $n$ integers (which can be chosen irreducible) so that 
\begin{equation}
\sum_{i=1}^{i=n} k_i \omega_i =0 \label{comcond}
\end{equation}
Such  a fulfilled condition (\ref{comcond}) reduces  the dimension $n$ of the corresponding invariant tori  by $1$. 

We now consider perturbations  of this integrable Hamiltonian 
\begin{equation}
H= H_0(\{I_i\}) + \epsilon h(\{I_i,\theta_i\}) \label{pertham}
\end{equation}
where $\epsilon$ is a small positive parameter and $h(\{I_i,\theta_i\})$ an arbitrary pertubative Hamiltonian which  depend both on the actions and the  angles. 
Then  we know from Poincar\'e, that in general all resonant tori  (\ref {comcond}) and a neighbourhood of them (resonance gap) are destroyed and replaced by many new trajectories  which could be either linearly unstable and chaotic or still   linearly stable and quasiperiodic.

However, it might be that for very special exceptional choice of the perturbation, for example if $h(\{I_i,\theta_i\})$ does not depend on the angles $\{\theta_i\}$  in some region of the phase space).  We have to say that the existence of chaotic resonance gaps is a generic property which means it might not be fulfilled in rare and special models.
Actually the mathematical definition of the word "generic" which could be found in the literature (e.g. a property valid in Baire subset), may  not be physically acceptable because it depends in which topological space we consider the problem. We  do not debate  this question.  We have to appreciate what means generic in the physical context we consider.

Despite the existence of infinitely many resonance gap,  all invariant tori  of the integrable Hamiltonian are not destroyed.   The frequencies of the non resonant (incommensurate) tori which surely survive to the perturbation  fulfill  a diophantine  condition  \cite{Poe01} that is there exists two positive constants $\alpha$ and $\tau>n-1$ such that for any set of integers $\{k_i \} \in \mathcal{Z}^n$ we have
\begin{equation}
|\sum_{i=1}^{i=n} k_i \omega_i |  \geq \frac{\alpha}{|k|^{\tau}} \label{diophante}
\end{equation}
where $  |k|=\sum |k_i|$.
It can be proven that the set incommensurate tori which fulfill this condition, has full measure in the phase space. 

Under smoothness conditions that the considered Hamiltonian is  an analytic function of its variables  (or if not at least $n-1$ differentiable with continuous derivatives), KAM theorem states that each of these incommensurate tori can continued up to some non vanishing critical value of the perturbation $\epsilon_c(\{\omega_i\})$ providing the Jacobian matrix $\{\frac{\partial \omega_i}{\partial I_j}\}= \partial^2 H_0(\{I_i\}) $ be invertible.  Moreover KAM theory, states that the global measure in the phase space of these surviving KAM tori  goes to full measure as the perturbation parameter $\epsilon$ goes to zero or equivalently the  measure of the whole set of resonance gaps shrinks to zero.

KAM theory can be used in many situations and especially in the vicinity of linearly stable periodic orbits (or just stable fixed point)  in any non integrable Hamiltonian.  It states as a corollary that the quasiperiodic solutions obtained within the linear approximation near  linear stable periodic orbits  survive as exact quasiperiodic solutions again on  non resonance conditions (\ref{diophante}).  Such quasiperiodic solutions not existing for the integrable Hamiltonian are called \textit{secondary} KAM tori while those existing at the integrable limit are called \textit{primary}. Finally, the  landscape of the trajectory behaviour in the phase space become complex  at all scale. At the largest scale, it consists of  primary KAM tori and their resonance gap which at a smaller scale contains secondary KAM tori  with their new resonance gaps  and so on... \cite{BC18} . The complementary part is mostly occupied by unstable chaotic trajectories. However, it has not yet been rigorously proven that they have a non vanishing measure though there are strong numerical evidences for that! 

\subsection{Arnold diffusion conjecture and extension}  \label{Arndiff}  A consequence of KAM theory is that the BEH hypothesis is not true  in principle for most Hamitonian systems because they  cannot spontaneously  thermalize because of the existence  many non ergodic KAM trajectories which occupies a non vanishing measure in the phase subspace $\mathbf{E}$ at constant energy$ E$.  Then it is natural to split  the constant energy subspace $\mathbf{E}=\mathbf{K} \cup \mathbf{C}$ into two disjoint complementary measurable parts. The set $\mathbf{K}$ consists of all KAM tori  which are linearly stable quasiperiodic trajectories. The complementary part $\mathbf{C}$  consists of the rest which are linearly unstable trajectories  mostly chaotic and ergodic (with strictly positive Lyapounov exponents). Many other non chaotic trajectories like periodic, quasiperiodic trajectories (such as Cantori),  whiskered tori, homoclinic and heteroclinic trajectories...  which all together have zero measure and can be neglected from the statistical point of view  (However they are important to study because kwowing more about them help to understand better the fine structure of  $\mathbf{C}$).  

The set   of KAM tori $\mathbf{K}$ is infinitely disconnected that is giving arbitrarily  two KAM tori and a continuous path connecting two points of them, in generic cases,  this path necessarily crosses resonance gaps which contains points not in $\mathbf{K}$. It is a fat Cantor set (fat because it has non vanishing measure). We say that  in $\mathbf{K}$ is \textit{porous} when it contains gaps (holes)  at all scale. It is the generic situation.

When the number of degrees of freedom $n$ is strictly larger than $2$,  KAM tori which have dimension $n$ cannot split  the $2n-1$ dimensional constant energy subspace $\mathbf{E}$   into two disconnected parts( unlike in the case $n=2$ where KAM tori have dimension $2$ with disconnected inside and an outside region).  Consequently when $n>2$ and when the complementary set $\mathbf{K}$ is porous, there are continuous paths in $\mathbf{C}$ connecting  two arbitrary points of  $\mathbf{C}$ so that  $\mathbf{C}$ is a connected set. This set is supposed to also have non vanishing measure  in $\mathbf{E}$  which is not equal to the full measure of $\mathbf{E}$ though it is nevertheless dense everywhere in $\mathbf{E}$. 

Denoting $\mu$, the Liouville measure restricted to $\mathbf{E}$  which is conserved by the Hamiltonian flow, this uniform measure may be written as the sum of  two complementary  measures $\mu = \mu_{\mathbf{K}}+\mu_{\mathbf{C}}$ defined for any measurable set $\mathbf{V}$ as  $\mu_{\mathbf{K}}(\mathbf{V})= \mu(\mathbf{V} \cap \mathbf{K})$  and $\mu_{\mathbf{C}}(\mathbf{V})= \mu(\mathbf{V} \cap \mathbf{C})$. Since both $\mathbf{K}$ and $\mathbf{C}$ are invariant subset under the Hamiltonian flow, measures $ \mu_{\mathbf{K}}$ and $\mu_{\mathbf{C}}$ are both conserved by this flow. Near the integrable limit KAM theory implies  that  $\mu_{\mathbf{K}} \approx \mu$ and $\mu_{\mathbf{C}} \approx 0$ while it becomes the reverse $\mu_{\mathbf{C}} \approx \mu$  and $\mu_{\mathbf{K}} \approx 0$ far from the integrable limit when most trajectories are chaotic.  

Arnold diffusion conjecture \cite{Arn64} claims  that when this set $\mathbf{C}$ is connected,($n>2$), there exists (in the generic case)  trajectories which can go arbitrarily close  to each KAM tori of  any arbitrarily given sequence of KAM tori.  Arnold gave a proof only in a special example, a generic proof was given only recently \cite{CX19}.  However it is not yet proven that most trajectories in $\mathbf{C}$ have same property.  We propose a more general conjecture which extends the Arnold diffusion conjecture,  we call  Boltzmann-Arnold  Ergodic Hypothesis because it is nothing but the Boltzmann hypothesis restricted to the only ergodic subspace $\mathbf{C}$  accessible to chaotic trajectories. 
 
 \subsubsection{Conjecture: Boltzmann-Arnold Ergodic hypothesis} \textit{When the number $n$ of degrees of freedom of the Hamilton is strictly larger than $2$, then in generic situations, the subset  $\mathbf{C}$ is ergodic that is most trajectories in $\mathbf{C}$ are dense everywhere in  $\mathbf{C}$ (and consequently in the whole subspace $\mathbf{E}$ of the phase space at constant energy $E$ ). Average in time of a physical quantity over a given trajectory is  equal  (with probability $1$) to the same quantity averaged over  $\mathbf{C}$ with its induced Liouville measure defined above as $\mu_{\mathbf{C}}$.}\\
   
Note that if the total measure of  the KAM tori would be zero $\mu_{\mathbf{K}}=0$,  the original  BEH would be recovered since we would have  then $\mu_{\mathbf{C}}= \mu$.  

This mathematical statement about the statistics for the distribution of states after a very long time, does not say anything about  the time needed for having this statistics. This hypothesis needs a physical interpretation because the subset $\mathbf{C}$ contains an infinite number of tiny resonance gap where according to the Nekhoroshev theorem \cite{Nek77}, the dynamics is almost the same as those of the neighbouring tori over very long time. Consequently diffusion through $\mathbf{C}$  becomes  very  slow in such regions  while it is much faster in regions with strong chaos. The escape time from such regions to fill densely other regions may become so long that their observation becomes numerically impossible. But note also that these tiny gaps where the trajectories could remain trap over very long time get negligible measure compared to the global measure of   $\mathbf{C}$. 

We do not have a rigorous proof of this modified conjecture but we have some arguments suggesting that chaotic and ergodic trajectories should fill a substantial part of the phase space we shortly describe:

\section{Dynamics of Hamiltonians on infinite  lattices} 

We  consider now wave packets in Hamiltonian on infinite square lattices at arbitrary finite dimension. 
If the system is spatially periodic, the linear spectrum is purely absolutely continuous with bands and delocalized eigen modes. When the system is random, the linear spectrum  may become purely discrete but in higher dimension than $2$, it may still contain an absolutely continuous part with mobility edges. 

When the linear spectrum contains an absolutely continuous part and when  a localized wave packet is introduced in the nonlinear system, the dynamics of the wave packet cannot be quasiperiodic neither chaotic because the Fourier spectrum of its  dynamics, which is dense necessarily involves  frequencies in the absolutely continuous part of the linear spectrum which radiates energy. However, the  wave packet  may generate a periodic solution if its frequency and all its harmonics do not overlap the absolutely continuous part of the linear spectrum. We may obtain stationary periodic solutions called Discrete Breathers (DB) \cite{MA94,Aub97,Aub07}. When DBs exist,  two well-known situations may occur for an arbitrary initial wave packet depending on its initial conditions. Either the initial wave packet spreads to zero or it converges to a Discrete Breather. In the later case, a part of its energy spread while the other part remains localized. Sometime we get a transiently mobile DB which finally stops as a stationary DB. These situations were studied elsewhere. 

The situation which was poorly understood, is the topics of this paper. It occurs when the linear spectrum is purely discrete without any continuous part that is when the linear spectrum is not dissipative \cite{AS09}.

\subsection{KAM tori in infinite Hamiltonian systems on lattices with linear discrete spectrum}

 A common belief \cite{FS75} is that as the number of degrees of freedom of a Hamiltonian  increases,  the relative measure of the KAM tori goes to zero. However this statement is quite imprecise. Actually, this statement is likely  true only in the thermodynamical limit, that is when the energy of the system is extensive (i.e. proportional to its size). This is not the situation here since the initial wave packet energy is finite.
 
A pioneering work by Fr\"ohlich, Spencer and Wayne (FSW)  \cite{FSW86} proved rigorously that Anderson localization may persist in the presence of nonlinearities. Their rigorous proof holds only in  some class models chosen in order  the complex KAM machinery be easier to implement. 
 
They consider a random system on a square  lattice at arbitrary dimension, with strong disorder where the eigenstates are mostly localized at single sites. They chose a quartic nonlinear perturbation which couples only nearest neighbour oscillators.$i:j$ denote the bonds of the lattice between site $i$ and its neighbouring site $j$ counted once since $i:j$ is equivalent to $j:i$.  The FSW Hamiltonian  has the form
 \begin{equation} 
 H_{FSW} = \sum_{i\in \mathcal{Z}^d} h_i(u_i,p_i) + \sum_{i:j} W_{i:j}(u_i,p_i,u_j,p_j) 
\label{FSW} 
\end{equation} 
The linear part   $\sum_i h_i(u_i,p_i)= \frac{1}{2} p_i^2 + \frac{1}{2} \omega_i^2 u_i^2$ of this Hamiltonian consists of a collection of harmonic oscillators representing the linear eigen modes of a model with Anderson localization. For simplicity, they are located at site $i$ of a $d$ dimension cubic lattice and their frequencies   $\omega_i$ are supposed to be random.  The higher order nonlinear terms $ W_{i:j}(u_i,p_i,u_j,p_j)$ are assumed to involve  only nearest neighbour oscillators $i:j$, and to be  analytic  functions with respect to all their variables with quartic expansion (that is without any term at order lower than $4$).  

\subsubsection{FSW theorem }   \textit{Choosing arbitrarily a strongly localized linear solutions (i.e. decaying faster than exponential see \cite{FSW86}) of the linearized Hamiltonian which thus is quasi periodic in time, the probability that there exists an exact quasiperiodic (non resonant solution)  of the perturbed Hamiltonian near this arbitrarily chosen linear solution, is non vanishing  when the amplitude of the initial solution becomes small enough. Moreover this probability goes to $1$ as this amplitude goes to zero. The perturbed solution is also quasi periodic with perturbed frequencies.}\\
(To be more precise,  the perturbed linear solution is continued by keeping constants  the initial actions $I_i$ of each unperturbed oscillator  $h_i(u_i,p_i)$.)

The probability of existence of these KAM solutions  is defined with respect to the random distribution of frequencies $\omega_i$ supposed to be uncorrelated and distributed according to some smooth probability law.  They also conjecture  that their theorem remains  still valid in general for lattice Hamiltonian where the linear spectrum exhibits complete Anderson localization  and when raising the fast decay conditions on the unperturbed   wave packet with finite energy. 

When the disorder of becomes weaker, the maximum localization length of the linear eigenmodes increases so that the nonlinear interactions between them should  extend significantly beyond nearest neighbours. Model (\ref{FSW}) should be improved by extending the  short range  interactions  beyond nearest neighbours.  These changes should  not change the FSW theorem about the existence of KAM tori.  But however, when the disorder becomes weaker,  the maximum linear localization length should increase and diverges (in more than 2 dimensional models, mobility edges should appear before disorder disapear). If the linear spectrum contains an absolutely continuous part,   KAM tori cannot exist anymore according to the  arguments presented at the beginning of this section. Consequently arguing the physical continuity of the dynamical behaviour of the model,  we expect that the small amplitude region where KAM tori become very dense, shrinks  to zero while  the longest localization length diverges. The consequence  is that it should be much easier to observe numerically KAM tori with significantly large amplitudes at stronger disorder than at weaker disorder. 
 
\subsection{Numerical observation of KAM tori in Large non Integrable Hamiltonian Systems}

The FSW theorem suggests that the existence of KAM tori as localized solutions in nonlinear infinite lattices with purely discrete spectrum is an ubiquitous phenomena. We tested an easy numerical method  in \cite{JKA10}  which does not require too much time consuming  calculations. It is based on the  theory  of Almost Periodic Functions pioneered  by Harald  Bohr in 1924 \cite{Boh}. 

\subsubsection{Bohr Definition and Theorem}  \textit{ A  function $f(t):\mathcal{R} \rightarrow \mathcal{C} $ is said to be almost periodic when for any $\epsilon >0$, there exists $L(\epsilon)>0 $ so that in any interval with length $L(\epsilon)$ there exists a pseudo period  $\tau$ so that $f(t+\tau)-f(t)| < \epsilon$ for all $t$}.

\textit{ If $f(t)$ is bounded uniformly continuous and  almost periodic, then $f(t)$ can be written as a generalized Fourier series $f(t)= \sum_n f_n e^{i \omega_n t}$ which is absolutely convergent and where $\{\omega_n\}$  is a countable set of frequencies.} \\

An almost periodic functions  generally  involves an infinite number of arbitrary fundamental frequencies unlike quasiperiodic functions which involve only a finite number of fundamental frequencies (and all their harmonics).  Thus the KAM tori in infinite systems correspond to almost periodic solutions while those in finite system are only quasi periodic. However we confuse often the two terminologies since it is not possible to discriminate numerically between the two. 

This Bohr theorem can be used to find KAM tori in large system by only testing Poincar\'e recurrence.  Poincar\'e recurrence theorem states that in a finite Hamiltonian system,  a given trajectory will  return with probability $1$ arbitrarily close to  its initial condition.   However it does not say anything about the recurrence time except that it is not infinite.

When a trajectory is almost periodic, Bohr theorem  states, that for a given accuracy $\epsilon$, the Poincar\'e  recurrence occurs repeatedly at  bounded intervals  of time which is rather easy to check numerically. On contrary  when the trajectory is chaotic, the Poincar\'e recurrence times are expected to be randomly distributed   and their intervals unbounded. Moreover, for chaotic trajectories which are ergodic in a substantial part  of the phase  space (on contrary to KAM tori ) the time lost to explore this ergodic domain makes that the average recurrence time  dramatically increases  as the number of degree of freedom increases.  As a result, Poincar\'e recurrence becomes practically not numerically  observable for chaotic trajectories even in relatively small systems.  However, it can be observed that non chaotic trajectories in large systems exhibits Bohr recurrence which then strongly suggests the existence of KAM tori. However, whatever is the required numerical accuracy, numerical observations cannot really distinguish  between true KAM tori or weakly chaotic trajectories arbitrarily close to them.  We can only say that the numerical observations are consistent with their existence as mathematics predict either in finite models and or in some infinite models \cite{FSW86} . 

We checked the existence of KAM tori  in 1D Random DNLS model for wave packet  initially localized  at a single site \cite{JKA10} that for a given disorder realization, and we indeed found  there are indeed for many wave packets mostly at small amplitude which fulfill the Bohr condition that is they are repeatedly recurrent  many times over  all the numerically observed evolution time.  Note that  it is sufficient to test the almost periodicity over only one coordinate located at the initial single site. Then if only one coordinate  of the trajectory is an almost periodic function of time, all other coordinates should be also almost periodic functions since they are related one with each others by the Hamilton equations. 

We also observed "sticking" trajectories which  look Bohr recurrent over a certain time beyond which recurrences completely stop. They can be interpreted as trajectories generated by initial conditions which are close to true KAM tori in narrow resonance gaps. As expected from the Nekhoroshev theorem  \cite{Nek77,BL20}, the perturbation growth near quasiperiodic integrable solutions is slower than any exponential but algebraic so that the trajectory may look a  KAM torus over a long time.  Actually, because of Arnold diffusion, trajectories in resonance gap nearby  true KAM tori  remains for a long time as very weakly chaotic trajectories confined within this gap. But finally they should escape from this resonance gap and rejoins the main chaotic region where it becomes strongly chaotic. Of course this escape time may become longer than the computing time so that it is not possible to really distinguish true KAM tory and weakly chaotic trajectories.  On the other side,  the width of the corresponding high order resonance gap also drastically goes to zero.   Thus in practice we observe that the region of KAM tori at small amplitude looks connected while we expect the existence of infinitely many tiny resonance gaps.  We also observed Bohr recurrent trajectories for large amplitude wave packets which may be interpreted by the fact the initial  condition is close to linearly  stable discrete breathers (which are known exist in that model at large enough amplitude). 

We also noticed that the trajectories which are not Bohr recurrent always exhibit a chaotic behaviour. We also checked that the trajectories looking KAM tori  according to this Bohr criteria have zero Lyapunov exponent  within the numerical errors while the all others  exhibit a chaotic trajectories with non zero Lyapounov exponent. Studying  special trajectories like  periodic cycles, homoclinic or heteroclinic trajectories, whiskered tori etc... which occupy a zero measure in the phase space in finite Hamiltonian, would require different  methods,  variational or else.  Actually, the Bohr recurrence method is a simple and efficient  method for discriminating numerically between  KAM trajectories and chaotic trajectories. 

It was also shown numerically with scaling arguments that  the probability that the wave packet looks regular when it is  spread over the whole system goes to unity as the size of the system diverge \cite{PS11}. Recently, we came aware of recent works \cite{SS22} which use the  different GALI method for discriminating between the chaotic trajectories  and the others (KAM tori) in a different model (random discrete KG models). Actually they got basically the same conclusion as ours in the random DNLS  \cite{JKA10} that is the probability to find regular trajectories (or KAM tori) goes to unity at small amplitude.  Discriminating between localized and spreading  chaos is another question discussed the next subsection. 

\section{Long time behaviour of a Wave Packets} 
We consider our infinite systems as usual in physics as a finite systems involving $N$ sites where $N$ becomes large. Though the existence of KAM tori proven by KAM theory for finite size systems is not questionable, the question is : would they disappear when $N\rightarrow +\infty$?  On the base of the above arguments, we assume that they do not and moreover remains very dense at small amplitude.  Then an  initial wave packet may generate a KAM torus with a finite probability in infinite system. It then  remains   stationary and quasiperiodic so that there is no spreading and no diffusion at all. An almost identical solution should also exist in finite size system much  larger than its spatial extension. 

On contrary, for chaotic wave packets, the extended Arnold diffusion conjecture described above should hold whatever is the size of the system. Then if one wait long enough, most chaotic trajectories should visit  the whole set of chaotic wave packets $\mathbf{C}$ according to the probability density  $\mu_{\mathbf{C}}$. At  small amplitude the phase space is mostly  occupied by  KAM tori so that chaotic wave packet have almost zero probability to exist. We propose  to quantify this intuitive effect into a measurable criteria which should help numerically to understand the wave packet behaviour at very long time.   

\subsection{A  Criteria for  the Long Time behaviour of a chaotic Wave packet}  

Assume first that a wave packet with energy $E$ is spread over the whole system (whatever is its dimensionality) which consists of large number of sites  $N$, its energy density per site is about its average $E/N$  and goes to zero as the system size $N$ grows to infinity.  Near this limit, the energy of the wave packet is mostly obtained from the harmonic terms in the Hamiltonian while the contribution of the higher order nonlinear terms to its  energy  nearly vanish. This situation of very weak nonlinearity is highly favorable  for generating KAM tori. Thus the measure of the trajectories which are both in  $\mathbf{C}$ and have a small amplitude, becomes negligible at the same time. On contrary, initial wave packets which are initially rather focused, have a probability to be in $\mathbf{C}$ which could reach unity.

Thus we need to use  a parameter  which measures how much  a given  wave packet is spread. The most commonly used criteria  is the  inverse participation number of the wave packet energy, but many other choice would be possible without changing  the physical interpretation.  To define a participation number, we split the global Hamiltonian $H$ of our system on a lattice into a sum $H=\sum_i H_i$ of local hamiltonians  $H_i$ at the lattice sites which can be chosen positive and vanishing when the corresponding oscillator is at rest.  We may choose for example, $H_i$ as the Hamiltonian of the isolated nonlinear oscillator at site $i$ plus half of its  interacting potentials with its neighbours. Then local energy of the wave packet at time $t$ and at site $i$ is  $H_i(t)$ and $E= \sum_i H_i(t)$ . With this definition,  the inverse participation number of a wave packet  becomes

\begin{equation}
S (t) = \frac{\sum_i H_i^2} {(\sum_i H_i)^2} =\frac{\sum_i H_i^2} {E^2} \label {invpart}
\end{equation}
In finite systems with size $N$ , we necessarily have $\frac{1}{N} \leq S \leq 1$ since the participation number cannot be larger than the size of the system. In the limit of infinite size $N$, this inequality becomes  $0 \leq S \leq 1$.

Then we can define a probability density $P(S)$ for arbitrary configurations in $\mathbf{C}$ with the fixed  energy $E$ to have a given participation number $S$. More precisely 
\begin{equation}
P(S) dS = \frac{\mu_{\mathbf{C}}(B(\left[S,S+dS\right]) } {\mu_{\mathbf{C}}(\mathbf{C})}
\label{probdens} 
\end{equation}
where $\mu_{\mathbf{C}} (B(\left[S,S+dS\right]))$ is  the measure  of the subset $\mathbf{B}(\left[S,S+dS\right])$  of wave packets in $\mathbf{C}$ which have  an inverse participation number  in the small interval $\left[S,S+dS\right]$ ( $\mu_{\mathbf{C}}$ is the  induced Liouville measure in $\mathbf{C}$ as defined above). 

If one assumes that there are no KAM tori, we have  $\mu_{\mathbf{C}}(\mathbf{C})=\mu(\mathbf{E})$, then the BEH would hold in the whole space $\mathbf{E}$ instead of $\mathbf{C}$. Thus the initial wave packet should spontaneously thermalize at zero temperature that is it should spread. In that case, instead of calculating $P(S)$ within the microcanonical ensemble,   it could be simpler to use the Boltzmann statistics in the grand canonical ensemble where the system is not strictly isolated. Then we have to fix a different temperature $k_BT_N=1/\beta_N$ for each systems with size $N$ in order that the  total energy average of the whole system $\mathbf{E}$ be equal to the energy of the wave packet.  Since  $e^{-\beta_N E_i}/Z(\beta_N) =\beta_N e^{-\beta_N E_i} $ is the Boltzmann probability energy density for each oscillator, we have $<E_i>=1/\beta_N$ and $k_BT_N=E/N$.The total energy $\sum_i E_i$ of this system is not  absolutely fixed as it should be but its expected fluctuation $<(\sum_i E_i- E)^2>^{1/2}= E/\sqrt{N}$  goes to zero  at large size $N$. We neglect this fluctuation. Then we can calculate the average of $S$ defined by (\ref{invpart}) which is  $<S> = \int P_0(S) S dS =S_m= 2/N$  and $<(S-S_m)^2>^{1/2} =\sqrt{20}/N$. This result shows that in the absence of KAM tori, $P_0(S)$ should exhibit a thin peak located very near zero which becomes a Dirac peak $\delta(S)$  in the limit of infinite size.

The real curve  $P(S)$  should be   drastically different from $P_0(S)$  because most initial wave packet  with a small participation number $S$ generates  KAM tori and has to be removed from the statistics. We expect that $P(S)$ should exhibit a broader peak located around an average participation number $S_m$ which is not vanishing. This peak roughly corresponds to the cross over energy amplitude $A_m  \approx E S_m$  below which  the density of KAM tori sharply increases (Note that because of spatiall disorder, this cross-over should depend on  the location of the wave packet in the system. $A_m$ is just an average over all the wave packet  location or equivalently over disorder). Thus when the wave packet  energy $E$ increases,  its peak at $S_m$ should move toward  smaller values  proportionally to $1/E$.   We also expect that $S_m$   becomes  smaller for smaller disorder  because when the maximum localization length increases the domain of existence of KAM tori shrinks to zero as explained above. 

\begin{figure}
\vspace{-1cm}
\includegraphics[height= 6cm,width=10cm,angle=-0]{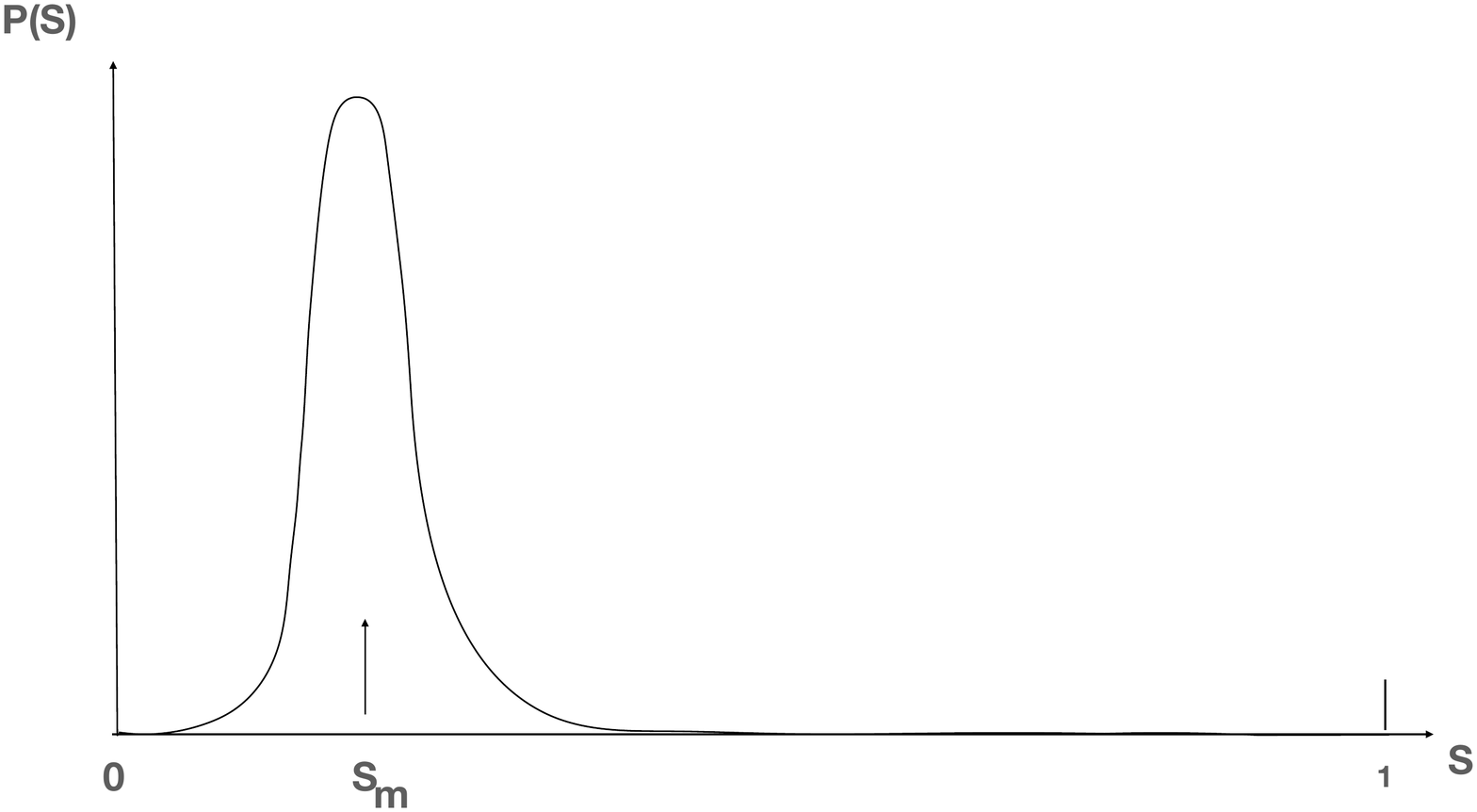}
\caption{Sketch of the expected probability density $P(S)$ for chaotic wave packets at the same energy   (in $\mathbf{C}$) versus their inverse participation number $S$ in the limit of large system} \label{fig1}
\end{figure}

We thus conjecture  that $P(S)$ should exhibit  a single peak as shown  figure (\ref{fig1}) with a maximum located at $S_m$ not near zero but somewhere else in the interval  $0< S_m <1$. 

This probability density $P(S)$ might be numerically calculable with a reasonable accuracy at least for small system. One method could be to make the statistics of the initial  inverse participation number $S$ over many  wave packets chosen  according to the Boltzmann probability in the microcanonical ensemble at the energy $E$ just discarding  all initial wave packets which look KAM tori (actually in finite systems KAM do exist). Another method would be to integrate one these initial chaotic wave packets over  very long time in order to visit the whole set $\mathbf{C}$ and to look at the distribution of  its participation number over time. According to the Boltzmann Arnold  hypothesis, the two methods should be equivalent. However it may be more efficient to combine the two methods by repeating the time statistics method  over many initial wave packets chosen randomly in $\mathbf{C}$. Increasing the size of the system as much as possible  could give a more precise idea of $P(S)$ at infinite size. Further work might be necessary for improving these methods for finding $P(S)$ accurately but nevertheless we believe that some numerical test on relatively small systems should give an idea of this curve.  Actually, up to now most numerical experiments were performed for well focused initial wave packets  located at a one or few sites with an inverse participation number a priori very different from the expected most probable value roughly around $S_m$ which should be obtained  the pseudo thermal equilibrium in $\mathbf{C}$ is reached.  If the initial wave packet is far from this equilibrium,  it should be necessary to wait a certain transient time (which may be very long) to reach the bulk of the subspace $\mathbf{C}$ where $P(S)$ is near its maximum and to then to see the  fluctuations of the wave packet participation number with a stationary statistics. 

\subsection{Long time behaviour of Wave Packet} 

The above studies suggest two essential conjectures about the long time behaviour of finite energy wave packets (not rigorously proven but supported by strong arguments):
\subsubsection{Conjecture 1: Absence of spreading}  \label{Conj1} \textit{ In an infinite  nonlinear lattice Hamiltonian system with a purely discrete linear spectrum (i.e. with Anderson localization), a wave packet (with finite energy) may generate:
\begin{itemize}
\item either  non spreading stationary KAM tori (in $\mathbf{K}$) with a finite probability 
which goes to unity  at small amplitude \\
\item or a chaotic trajectory (in $\mathbf{C}$) with the complementary   probability. Moreover the probability that  such  chaotic wave packet 
spreads (i.e. its maximum amplitude goes to zero), is zero. 
\end{itemize}}

In physical terms it just never spread. Consequently at any given time the wave packet remains  spatially localized in the sense that its inverse participation number is  non vanishing.  However this condition does not implies that  the wave packet always remain well focused with a single peak. When it is chaotic it may also  break into a finite number of smaller  peaks moving more or less independantly.

Otherwise, in a finite system, the Boltzmann Arnold hypothesis  implies  that any initial wave packet  with chaotic behaviour should go at some time arbitrarily close to any other chaotic  wave packet  in $\mathbf{C}$ with the same energy at any other spatial location.  This property  implies energy diffusion through the system (but no spreading)  in  any finite systems whatever is its size and consequently  in the infinite system as well.

\subsubsection{Conjecture 2: Diffusion} \label{Conj2}  \textit{ With the same conditions as for the above conjecture, the probability that  a chaotic wave packet (in $\mathbf{C}$)  be spatially diffusive (i.e. visit any spatial location in the lattice)  is unity }\\

These statistical arguments say nothing about the dynamics of the wave packet diffusion. 
Actually we expect that the rate $D$ of diffusion  of the energy of the wave packet   depends both on its inverse participation number $S$ and its spatial location which both fluctuates as a function of time. When   $S$  is larger than or of the order of $S_m$ , the wave packet is strongly chaotic  because KAM tori if any are relatively rare, we should get rather fast diffusion.  On the other side  when $S$  becomes smaller than or of the order of $S_m$,  the KAM tori become denser,  so that chaos becomes weaker  and diffusion slower.  It may also happen at smaller $S$ that wave packet get trapped for a very long time in very weakly chaotic  in tiny resonances gaps.   Because of Nekhoroshev theorem,  the escape time from  such quasi trapping region increases very fast by order of magnitudes as $S$  becomes  small and there is practically no diffusion during this trapping time.
However, though  quasi trapping  events last very long, they are also very rare so that the probability density $P(S)$
remains small for small $S$.  Thus, the  diffusion of the wave packet  may be sometime  fast , slow or very slow  depending not only on the variation of $S$ but also on the local disorder. Such a behaviour is reminiscent of a so call random walk in a random scenery
or of diffusion in porous media.  This fact may explain why standard energy diffusion is not observed but slower diffusion with lower time exponent in many models both with  disorder and nonlinearities.

This second conjecture is only generic. It may not hold exceptionally in some special models because it is a consequence of the Boltzmann Arnold diffusion  conjecture in $\mathbf{C}$   which relies on  the porosity of the set $\mathbf{K}$  of KAM tori at constant energy $E$  at the same energy $E$ as explained in subsection (\ref{Arndiff}). 
In some special  models,  there may exist a   region $\mathbf{R}$ of the phase space where regular KAM trajectories do exist  but the expected resonance gaps in the neighbourhood of each resonant tori  at the integrable limit do not open. The  consequence is that  the subset $\mathbf{K}_{R}=\mathbf{R} \subseteq \mathbf{K}$ of  KAM tori 
 is not porous but remains  dense and connected. In  that  region  the nonlinear perturbation of the Hamiltonian preserves complete integrability  (we call such an Hamiltonian semi-integrable) and  consequently the set $\mathbf{C}$ of chaotic trajectories cannot be dense in $\mathbf{K}_{R}$ and thus not everywhere in $\mathbf{E}$ as it should. Moreover $\mathbf{C}$ may become disconnected because the proof of connectivity of $\mathbf{C}$ also relies on porosity. Full ergodicity in $\mathbf{C}$ is not true but however, ergodicity may persists in each of the connected subset $\mathbf{C}_n$ separately.  The Ding Dong model described in the next is an example of such non generic model.

\section{Exact results for Ding Dong models} 

The class of Ding Dong  (DD) models we discuss now are example of such non generic models  (where subdiffusion was nevertheless observed  in some 1D cases) by A. Pikovsky  \cite{Pik20}). We can prove that they are semi integrable at small enough amplitude as defined above. More exact results can be obtained  on this class of models. We prove the  first conjecture (\ref{Conj1}) of the previous  section that no wave packet  in DD models can spread to zero  We also prove that the set $\mathbf{C}$ of chaotic trajectories is disconnected (unlike the generic case). As a consequence,  the second conjecture (\ref{Conj2}) does not hold for DD models due to the absence of porosity of the regular (i.e. KAM) trajectories. The consequence of the existence of non porous  \textit{KAM barriers} makes that localized chaotic wave packets may exist unlike for the generic cases. We also explain how  perturbations of this model restore its genericity.

\subsection{DD model Definition}

The hamiltonian form of the Ding Dong (DD) model is those of a modified FSW model where  the nonlinear coupling is replaced by hard core potentials.  It consists  of an array at arbitrary dimension of random anharmonic oscillators on a lattice 
at arbitrary dimension $d$  with nearest neighbour hardcore coupling. We choose square lattices for simplicity but our results would hold for any network even random on condition that  the coordination (the number of bonds starting from a given site)
be finite and bounded. 
\begin{equation}
\mathcal{H}_{DD} = \sum_i  (\frac{1}{2 m_i} p_i^2 + V_i(u_i) )+ \sum_{i : j} W( u_j+b_{i:j}-u_i)
\label{HDD} 
\end{equation} 
where the smooth anharmonic local potentials $V_i(x)$  are chosen randomly and  expands as  $V_i(u_i) \approx   \frac{1}{2}m_i \omega_i^2+ ...$.  We assume the existence of two positive constant $K_m< K_M$ so that  
\begin{equation}
 K_m x^2 \leq  V_i(x)< K_M x^2\quad \mbox{ for all }  i   \quad \mbox{and}  \quad x
 \label{vbond}
 \end{equation} 
 so that the lowest order of the expansion never  vanishes.
The masses $m_i$ and linear frequencies $\omega_i$ are random uncorrelated random numbers  with smooth probability density bounded in  intervals  $0 < m_{min} \leq  m_i \leq m_{max}$ and $0< \omega_{min} \leq  \omega_i \leq \omega_{max}$. 

The interaction potential $W(x)$  is a hardcore potential  
\begin{eqnarray}
W (x) &=& 0 \quad \mbox { for } \quad  0 \leq x
 \nonumber \\
W(x) &=& + \infty  \quad \mbox {for }   x < 0 
\label{hardcore} 
\end{eqnarray}
 $\{b_{i : j}\}$ is a given collection of  positive numbers attached to each bond $i:j$ between nearest neighbour oscillators $i$ and $j$.  These numbers are assumed to be randomly distributed in a finite positive  interval  
 $\left[b_{min},b_{max}\right]$
 so that 
 \begin{equation}
 0< b_{min}<b_{i : j} \leq b_{max}
 \label{bimin}
 \end{equation}   
We may assume  $b_{min}=0$ but on condition the probability density of $b_{i : j}$ be not singular near zero, more precisely the probability that $b_{i : j}$ belongs to intervals $\left[0,\alpha \right]$ near zero, goes to zero as the width   $\alpha$ of this interval goes to zero. For more generality, we consider this situation $b_{min}=0$ in the next.   
This condition has the advantage to allow the existence of chaotic trajectories in the model even at very small energies similarly to  the other generic models considered in this paper. 
 
Consequently, the algebraic distance between two neighbouring oscillators $d_{i:j}(t) = u_j(t) - u_i(t) + b_{i : j} $ has to remain positive or zero for all trajectories. During some interval of time between two collisions, each  oscillator $i$  has a time constant energy $E_i=\frac{1}{2 m_i} p_i^2 + V_i(u_i)\geq 0$  and oscillates periodically with a frequency $\Omega_i(E_i)$  which generally depends on its energy (except if we choose the potential  harmonic). When  a collision occurs within  one bond $i:j$ 
among the $2d$ bonds connected to $i$ (that is when $d_{i:j}(t)$ vanishes at some time $t_c$),  a standard elastic collision 
occurs between the two neighbouring oscillators $i$ and $j$ where its energy (and frequency) changes  discontinuously according to the conservation laws  of the bond energy $E_i+E_j$ and  momenta $p_i+p_j$ at the collision time. 

\subsection{Regular trajectories and Lemma}
Exact result about (non) wave packet spreading are rather easy to obtain on this class of model  as a consequence of the following lemma which is straightforward to prove:
\subsubsection{Lemma} \textit{Considering a bond $i:j$ assumed isolated from the rest of the system, the two oscillators $i$ and $j$ cannot collide if
$E_i+E_j < B_{i:j}$  where $B_{i:j}= \min_x (V_i(x) +V_j(b_{i:j} -x)) $}\\

Because of the inequality eq.\ref{vbond},  we have 
\begin{eqnarray}
 \min_x  K_m(x^2+(b_{i:j} -x)^2 ) &=& \frac{K_m}{2} b_{i : j}^2<\min_x (V_i(x) +V_j(b_{i:j} -x)) = B_{i:j}  \nonumber \\
& < &\min_x  K_M(x^2+(b_{i:j} -x)^2 )= \frac{K_M}{2} b_{i : j}^2 \label{bbound}
\end{eqnarray} 
so that the collection of random numbers $B_{i:j} $ is bounded from below by  $ \frac{K_m}{2} b_{i : j}^2$ and from above 
by  $ \frac{K_M}{2} b_{i : j}^2$ since $b_{min}=0< b_{i:j} <b_{max}$.  

This lemma is also a necessary condition for having a pair of non colliding oscillator with the extra condition that is the frequencies $\Omega_i(E_i)$ and $\Omega_j(E_j)$ of these two oscillators be incommensurate number that is $n_i \Omega_i (E_i)+n_j\Omega_j(E_j) \neq 0 $  for any choice of integers $n_i$ and $n_j$. Otherwise, there may exist non colliding solutions at larger energy for example when the two oscillators $i$ and $j$ oscillates at the same frequency and in phase. However such situations have probability zero to occur in the whole phase space of initial conditions
if one assume that the set of frequencies $\Omega_i(E_i)$ are not constant but do depends on $E_i$. 
 
Considering now the infinite system, we then get a collection of random strictly positive numbers  $B_{i:j}$ attached to each  bond $i:j$. These random numbers only correlated between  nearest neighbour bonds $i:j$ and $i:k$ or $k:i$ sharing a common lattice site. 
As a corollary of this lemma, we obtain

\subsubsection{Corollary: Existence of regular trajectories} \textit{The  trajectories generated by the DD Hamiltonian (\ref{HDD}) which fulfill the set of inequalities 
\begin{equation}
E_i+E_j \leq B_{i:j} \quad \mbox{for all bond} \quad i:j
\label{KAMc} 
\end{equation} 
 at some arbitrary time, cannot exhibit any collision  at any bond and  any time, past and future.  }\\

Thus, when ineq.\ref{KAMc} is fulfilled for some initial condition,  each oscillator $i$ is oscillating periodically independently one from the other with its own period $\Omega_i (E_i)$,
and thus the global solution  is  almost periodic (regular). We consider them as (trivial) KAM tori since globally they are almost periodic  solutions. Thus   considering  the phase subspace $\{p_i,u_i\}$ fulfilling conditions $E_i+E_j < B_{i:j}$  for all $i:j$ 
 determines a bounded closed (compact) set $\mathbf{R}$ which contains $\{0\}$. 
 DD models are semi integrable according to the above definition  because  in that domain $\mathbf{R}= \mathbf{K}$, none of the  many resonant tori  (which are dense in $\mathbf{R}$) are destroyed by the nonlinear coupling   and of course the associated resonance gaps generating chaotic trajectories do not open as well. Consequently, the set of KAM $\mathbf{K}$ tori  remains connected and is not porous conversely to  the generic cases.
Consequently,  the complementary set of chaotic trajectories $\mathbf{C}$ in $\mathbf{E}$ is not connected and  the Boltzmann  Arnold diffusion conjecture cannot hold in the whole subspace $\mathbf{C}$ then allowing localized chaos to exist  as proven in the next. 

Condition (\ref{KAMc} ) is also generally necessary for having no collision so that if ineq.\ref{KAMc} is not fulfilled for only one bond $i:j$ (or more), the corresponding trajectory is generally chaotic with probability $1$. Then it is straightforward to prove that if a trajectory exhibits a single collision, it must exhibit an infinite number of further collisions ( not necessarily at the same bonds) as well in the past and as in the future. Consequently, the two sets $\mathbf{R} = \mathbf{K}$ are equivalent  neglecting only a zero measure subset. 

\subsection{Exact results} 
Using the lemma and ineq.\ref{bbound} , we readily prove:

\subsubsection{Theorem 1: Non spreading theorem} \textit{ No wave packet in a DD model  (\ref{HDD}) can spread.}\\

\textit{Proof:}  Let us assume that we find a wave packet with finite energy which would spread. Then for any $\epsilon>0$,
there should exist a time $t(\epsilon)$  so that  $0<E_i(t)<\epsilon$ for all $i$ and all time $t>t(\epsilon)$. 

The two nearest neighbour oscillators $i$ and $j$ connected by  bond $i:j$ cannot collide while  ineq.\ref{KAMc}  is fulfilled
and thus cannot exchange  any energy unless one of the  energies either $E_i$ or $E_j$ change
(because of a collision at an adjacent bond). This bond  is called blocked during the time where ineq.\ref{KAMc}  is fulfilled. 
When $\epsilon$ is chosen small enough, the local energies $E_i$ become small enough so there are many bonds  which become blocked for all time $t>t(\epsilon)$. 
In the simpler case where the colliding distances do not vanish that is  $b_{min}\neq 0$ in ineq.\ref{bimin},   eq. (\ref{bbound}) yields $B_{i:j} > B_{min}= \frac{K}{2} b_{min}^2 >0$ for all bonds. Then  assuming wave packet spreading, we can choose $\epsilon<B_{min}/2$ so that all bonds become  blocked forever after time $t(\epsilon)$. Since no more energy transfer between any oscillators can occur, spreading necessarily stops (actually, it should stop before this situation occurs). This contradiction proves the non spreading theorem. 

When the colliding distances are not bounded from below that is when  $b_{min}=0$, the theorem remains valid but its proof requires to use percolation theory which is well-known in physics.  The reader may find  many textbooks  in the literature defining this concept and moreover more exact results not useful  at the present stage of our study. 
Assuming the wave packet spreads to zero, when decreasing  the value of $\epsilon$, the density $D_b(\epsilon)$ of blocked bonds increases but reaches $1$ only at the limit $\epsilon=0$  (on condition  the density probability of the parameters $b_{i:j}$  be not singular as specified in the DD model definition). Simultaneously,  the density $D_u(\epsilon)= 1-D_b(\epsilon)$ of unblocked bonds decreases to zero so that it should reach the percolation threshold at $D_b^{\star}$ at some non vanishing $\epsilon_p$. Thus, when $\epsilon<  \epsilon_p$, the  set of unblocked bonds consists of  finite size connected clusters separate one from each other by blocked bonds. The consequence is that  the energy of the wave packet confined in each of these  finite cluster must remain constant for $t>t(\epsilon_p)$ since no energy transfer is possible through blocked bonds. 
Consequently spreading should stop  before $\epsilon$ reach zero which contradict the initial assumption and prove the non spreading theorem.
Note that in 1D model, the percolation threshold is obtained as soon there is a blocked bond that is when $2 \epsilon < \sup_{i:j} B_{i:j}$

\subsubsection{Note about the proof}  For defining a percolation threshold, we assumed that the random numbers $B_{i:j}$ in (\ref{KAMc}) are uncorrelated one with each other which is usually assumed for percolation theory but is not  true in our case (for adjacent bonds). For being absolutely rigorous, we may consider a subset of the set of blocked bonds defined  by  the condition  $E_i+E_j< B_{i:j}^{\star}=\frac{K}{2} b_{i : j}^2$  ( which implies (\ref{KAMc})). We  gain that now $B_{i:j}^{\star}$ are uncorrelated random numbers unlike $B_{i:j}$. The complementary set of bonds defined by the reverse  condition 
$E_i+E_j >B_{i:j}^{\star}$ contains the set of unblocked bond. Then decreasing $\epsilon$ below some non vanishing critical value $0<\epsilon_p^{\star}$,  this larger set also percolates  and then consisting of disconnected finite clusters.  Consequently the original smaller set of unblocked bonds is also disconnected which implies that it already necessarily exhibited  a percolation transition  at  some critical value of $ \epsilon_p>\epsilon_p^{\star}$ so that the above proof which requires that  $\epsilon_p$ is non vanishing still hold.

\subsubsection{Theorem 2: Localized chaos} \textit{For any choice of the initial energy $E$,  there exists initial wave packets at this energy which generate 
either quasiperiodic trajectories, or localized chaos (which may consist of one or several stationary chaotic spots).   This result implies that the set $\mathbf{C}$ of chaotic trajectories is never  fully connected at any energy $E$ unlike the generic models where Arnold diffusion conjecture hold.  }\\

\textit{Proof}: The proof of this theorem also use percolation theory as above. The bonds $i:j$ (if any) where $ E < B_{i:j}/2$ are blocked bonds (as defined just above) at any time since $E_i+E_j < 2E < B_{i:j}$.

We  first consider small amplitude wave packets with initial energies $E< E^{\star} $ not too large  so that the density of unblocked bonds $D_u(E)=1-D_b{E} $ be smaller than its percolation threshold $D_u(E^{\star}) =D_u^{\star}$ defined in the previous proof just above. Then considering an initial wave packet with energy $E< E^{\star}$, with initial energy 
arbitrarily distributed among the bonds of one of the many possible  connected cluster of unblocked bond at the same energy $E$, the energy of corresponding trajectory generated by this wave packet will  never escape from this cluster. Either it generates a regular quasiperiodic trajectory or it  generate a chaotic trajectory if there is at least one colliding bond. For having colliding bonds it suffices to choose  the distribution of the initial energy focused on the two sites of  an unblocked bond $i:j$ where $B_{i:j} < E$. Since $\min B_{i:j} =0$ , it is always possible to find such a bond.
Then the corresponding trajectory will be chaotic and remains confine in the cluster to which belongs the initiallly excited bond. 

When the initial energy  $E$ is larger than $E^{\star} $, one can still built such stationary solutions but the wave packet should not be  initially well focused but initially broken  into several smaller wave packets focused at different locations.For that purpose, we split arbitrarily its initial energy as a finite sum 
$E= \sum_{\alpha=1}^s E_{\alpha}$ where $0< E_{\alpha} < E^{\star}$  and $s$ is the number of "fragments".  
Then using the previous method for each fragment, we can construct confined  solutions with energies $E_{\alpha}$ located in different clusters of unblocked bonds which can be   arbitrary chosen. Each of these solutions may be chosen quasiperiodic or chaotic. Since the clusters are separated by blocked bonds, there is no energy transfer between these clusters, so that the dynamics in each cluster remains independent from each other. No energy diffusion is possible outside the clusters. 
This result proves that  the  phase space $\mathbf{C}$ is highly disconnected at any energy $E$ since we can built infinitely many different chaotic solutions at the same energy  which are localized and stationary at arbitrary but different locations. Thus  $\mathbf{C}$ should contain infinitely many disconnected parts $C_n$.

 However we may believe that when the  wave packet is initially well focused with an energy much larger $E$  than $E^{\star}$ ,  such  wave packet has almost full probability to generates delocalized chaos  though without spreading but with energy diffusion. We have not proof for this statement  but it is reasonable  to assume it for understanding why subdiffusion has been observed at least in some DD models \cite{Pik20}. 

Other exact statements can be easily proven using the above lemma:
\subsubsection{Theorem 3} \textit{Let us consider a DD model (\ref{HDD}) at some non vanishing temperature $k_BT=1/\beta$ , and 
a random configuration chosen according to the Boltzmann statistics, then the probability that this configuration be quasi periodic is zero.} \\

This theorem implies that the existence of regular trajectories does not plat any role for the thermal equilibrium. It confirms that at finite temperature most trajectories  are colliding solutions  (likely chaotic) so that the Boltzmann Ergodic Hypothesis  which is fundamental for statistical mechanics is true and that at finite temperature we have spontaneous Boltzmann thermalization in DD models but this is not the situation we consider in this paper.
Another theorem can be proven in DD models for confirming that the dynamical behaviour of a finite energy and chaotic wave
packet cannot be described according to the Boltzmann statistics in the full phase space, but within a statistics restricted within  each of the ergodic connected subsets $\mathbf{C}_n$ the existence was proven above . 

\subsubsection{Theorem 4} \textit{Let us consider a sequence of finite size $N$ DD models at non vanishing temperature $k_BT_N=1/\beta_N$, chosen in order that the average total energy $E$ (finite and  non vanishing) of each system is the same and equal to $E$ independently of $N$, Then the probability $Q_N$ according to the Boltzmann statistics that a configuration generates a quasi periodic trajectory, goes to unity when $N$ goes to infinity.}\\

\subsection{Non genericity of DD models } 

We proved that our first conjecture (\ref{Conj1} that wave packets does not spread to zero is true in this class of DD models but
our second conjecture (\ref{Conj2} has been proved to be wrong. Moreover, even the rigorous results  \cite{CX19}  giving a partial proof 
of the Arnold diffusion conjecture (as a generic property of Hamiltonian systems) becomes wrong in DD models. We discuss now this apparent contradiction  as  a consequence of the non genericity of the DD model.
Actually, if a  property proven for generic Hamiltonians,  is not fulfilled for a particular Hamiltonian, this property can always  be restored by adding  some arbitrary small perturbation to the non generic  Hamiltonian.  There is apparently a very sharp discontinuity between two very different qualitative behaviours between two Hamiltonians which are almost identical. This fact requires a physical interpretation since we should expect a kind of physical continuity.  

For recovering genericity, it suffices to  replace  the singular  hardcore potential $W(x)$ (\ref{hardcore}) by a smooth potential in the Hamiltonian (\ref{HDD}) which looks almost the same. Many choice are possibl, for example a simple choice
$W_{\nu} (x)=  e^{\nu x} $ with $\nu>0$ large so that $\lim_{\nu \rightarrow+\infty} W_{\nu} (x) =W(x)$ becomes the hard core potential of the original DD model.  
We may prefer to choose a potential which depends of the bond $i:j$ so that 
$W(u_j-u_i -b_{i:j})$ is replaced by  $W_{i:j,\nu}= (u_j-u_i)^4   e^{\nu (u_j-u_i -b_{i:j})} $
so that the perturbation be quartic at lowest order. Choosing $V_i(x) \approx \frac{1}{2} \omega_i^2 x^2+ a_i x^4+..$  we recover 
a FSW Hamiltonian (\ref{FSW}) were the existence proof of KAM tori hold in the infinite system. 
With such kind of perturbation with  $\nu$ large , the semi integrabiiity of DD model should disappear that is there are resonance gaps  which open near each resonant torus in $\mathbf{K}$. 
This feature sharply changes the qualitative behavior of the model. Considering wave packets in the original DD model with energy $E$ located at a  blocked bond $i:j$ characterized by $B_{i:j}>E=E_i+E_j$, we obtain only  non colliding quasiperiodic (KAM) solutions at energy $E$. We 
have shown that this  fact is at the origin of  the mechanism which blocks energy transfer  through the bond and allow the possibility of localized chaos.  In the slightly modified DD model, most quasi periodic KAM tori  located at the blocked bond survives and looks   almost the same but  then infinitely many very  tiny chaotic resonance gaps open near each resonant tori. Consequently  very weakly chaotic wave packets at  energy $E$ should also exist within  this   DD barrier initially impenetrable. Though they have a very small ( but non vanishing) probability to be found,  they  allow a slow energy transfer 
through the bond so that we can say the DD barrier become slowly leaking because of the small perturbation.

The consequence is that the subspace of chaotic trajectories  $\mathbf{C}$ recovers (in principle)  its connectivity so that our second conjecture should be true. However with such very small perturbation of the DD model  it should need a tremendously long time to see that blocked bond are slowly leaking so that localized chaos cannot  survive in general.
Nevertheless, the DD models reveal diffusion barrier which should survive transiently in generic models because of the expected local fluctuation of  
the density KAM tori. We conjectured above that these fluctuatiion should explain subdiffusion in the generic case.

\section{Consequences for Numerical Observations and Discussion} 

Our predictions have been developed  essentially to be consistent with the known mathematical theorems about KAM theory and the Arnold diffusion conjecture ( partially proved \cite{CX19}). Note also that they are strongly supported by exact results on Ding Dong models despite it is not  a generic model. Thus we have only qualitative predictions without precise quantitative informations. 

Some of our conclusions agree with the numerical observations which were done up to now while for others, there is no apparent agreement
( we shall explain why within our theory).  
We emphasize that our predictions are only valid for finite energy wave packets in infinite nonlinear Hamiltonian lattice models at any dimension but  with purely  discrete linear spectrum
(and does not hold for example for FPU models which have an absolutely continuous linear spectrum). 
We discuss now our main qualitative predictions versus numerical observations.

\begin{itemize}

\item  [1-]  Our first claim is that  initial wave packet may generate stationary quasi periodic KAM solutions with some non vanishing probability which increases and go to unity at small amplitude. Such trajectories are non chaotic and look very similar to those obtained in the purely linear case. Despite their existence as exact solutions was not  believed to be important for the wave packet diffusion problem, pioneering works \cite{FSW86} have proven their existence in some models. Otherwise  our early numerics deeply supported their existence  \cite{JKA10} in the random DNLS model and were also supported by scaling arguments associated to numerical simulations in \cite{PS11}. More recently such regular trajectories were observed again in different models with the different GALI  method \cite{SS22}
and confirm their existence as "regular trajectories". Their existence is indeed essential  for understanding the behaviour of chaotic subdiffusive wave packets.

\item [2-] For the other wave packets which do not generate KAM tori (with the complementary   non vanishing probability), most of them are chaotic which we is  confirmed by  numerical observations. Moreover, as a consequence of the Arnold diffusion conjecture, they should exhibit subdiffusion characterized by the divergence of the second momentum of the energy distribution as a function of time occurs. Note however that in non generic models like Ding Dong models,  some chaotic wave packets may remain spatially localized  at stationary chaotic spots and non diffusive. This feature is a consequence of  the (non generic) semi integrabilty of the Ding Dong model as explained at the end of the previous section. 
Moreover we suggested arguments that in generic cases, this wave packet diffusion may be similar to a random walk in a random scenery ( or diffusion in porous media) 
which could an interesting clue for explaining why   the observed diffusion exponent is different and smaller than for standard diffusion. 

Note that  impassable barriers  were clearly indentified in the DD model (allowing stationary localised chaotic wave packet ) which becomes slightly  permeable when genericity is restored by a weak perturbation. Then the diffusion of the wave packet should tremendously slow down when passing such quasi barriers.

\item [3-]  As consequence of the existence of KAM tori  which get full density at small amplitude, chaotic  wave packets cannot spread to zero. The explanation is an entropic effect due to the fact he phase space is almost  fully occupied by these KAM tori at small amplitude so that the wave packet has to remain in the available portion of the phase space at larger amplitude. Non spreading is  equivalent to say the inverse participation number  $S(t)$ does not go to zero. We predict  it fluctuates around an average value $S_m$  (see fig.\ref{fig1}) since it cannot go to zero. Though this statement  may be still considered as a conjecture in general, it is rigorously proven to be true in the Ding Dong models where numerical observations done for some of them  in 1D exhibits  wave packet subdiffusion similar to those observed in the other nonlinear random models \cite{Pik20}. 

\item [4] On contrary, it has been claimed on the base of numerical observations \cite{ASBF17}, that the inverse participation number goes to zero with an exponent $\sigma$ related to the subdiffusion exponent $\alpha$. We disagree with their interpretation.

\item  [5-]  Indeed, it has been recently observed on the base of numerical observations \cite{SGF22} that strong chaos persists in the wave packet dynamics after a long time.  This result  cannot be explained if the wave packet do spread since at small amplitude the nonlinear terms in the Hamilton equations should become very small. Consequently  over short time scale the dynamics should be almost the same as in the purely linear case.  Chaos could be observed only  after a long enough time scale but only as weak chaos of  Nekhoroshev type. On contrary, if one believe  that the wave packet does not  spread to zero,  its inverse participation number cannot vanish but only fluctuate around an  non vanishing average so that chaos  necessarily located at one of few chaotic spots could persist at long time  as it is observed.

\item  [6-]  A possibility for the long time behavior of a wave packet which has been suggested in the literature is that an initially chaotic wave may become regular after a long time and non chaotic which we interpret as being asymptotic to some KAM tori.   Actually this event has probability zero to occur as a straightforward consequence of  the Poincar\'e recurrence theorem. This theorem  states that for  measure preserving 
 dynamical systems (thus including Hamiltonian systems),  the probability that a trajectory be recurrent is unity. Since by definition trajectories  asymptotic to quasiperiodic KAM solutions (or any other solutions) are non recurrent, they have zero probability to occur.

\item [7-]  Finally we predict also the existence of a transient regime when the wave packet is initially well focused 
with an inverse  participation number near unity  far from its average equilibrium value $S_m$. The wave packet is generally  initially  far from its equilibrium  in $\mathbf{C}$ with the inverse inverse participation number $S\approx 1$ where $P(S)$ is almost negligible (see \ref{fig1}. Then it should first relax toward  the zone of the phase space corresponding to the bulk of $\mathbf{C}$ ("maximum entropy") corresponding to the peak of $P(s)$. During this transient time,  the wave packet should exhibit  simultaneously  diffusion and spreading that is both  the second momentum of the energy distribution  and its participation number $1/S$ grow. After this transient time only the second momentum continue to grow while the  inverse participation number stops to decay  and  only fluctuate around its well-defined average value $S_m$ shown at the maximum of curve (\ref{fig1}).  
Up to now this transient regime has not been numerically identified but it might have been a bias for correct  interpretations of early numerical observations. 

Note that this relaxation time is expected to become very long when the average inverse participation number $S_m$ is small. This situation occurs when the wave packet is initially well focused around a  single site with $S\approx 1$ with a  large initial energy and/or when the disorder is weak so that  the maximum linear localization length become large. In the limit of infinite localization length, KAM should completely disappear  so that
$P(S)$ becomes a Dirac peak at $S_m=0$. In that limit only, we should see complete wave packet spreading. 

\item[8-]  We also explained  that localized chaos does not generically exist because  the set $\mathbf{K}$ of KAM tori is generically porous which allows Arnold diffusion. However, we also rigorously proved that localized chaos do exist in (non generic)  Ding Dong  models where the set of KAM tori in non porous
and we also explain that this effect should disappear under arbitrarily small perturbations which restore porosity of the KAM set $\mathbf{K}$.
However localized chaos may be numerically observed over very long time (but not forever)
so that it looks stationary but on condition the average participation number $S_m<1$ be relatively large that is comparable to unity. This situation rather occurs in models near the limit opposite to those required for  having long transient we just described in the previous item, that when the  initial energy of the wave packet is rather  small and/or and  when the disorder is strong (so that the linear localization length is short). The reason is that the density of KAM tori then becomes much larger
so that the probability to find (weakly leaking) KAM barrier ( reminiscent of the blocked bond in  DD models described in the previous section) becomes much higher.
\end{itemize}

\section{Concluding Remarks} 

The cause of the absence of spreading of wave packets  in the class of models we considered here is essentially an entropic effect, that is the accessible volume in the phase space  which corresponds to very spread chaotic wave  packets becomes negligible compared to the global accessible volume for the chaotic wave packets. 
It is different from the absence of spreading which may occur in model which have an extra time invariant (or more) beside the total energy where the absence of spreading is due to topological constraints. Examples are DNLS models (random or not) which conserve both energy and total norm. In such model any trajectories must stay both in the $(2n-1)$ dimension manifold $\textbf{E}$ corresponding to its constant energy $E$ and in the $(2n-1)$ dimension manifold $\textbf{N}$ corresponding to its constant initial norm $N$. When the ratio of the energy $E/N$  is too large it turns that 
the manifold $\textbf{E} \cap \textbf{N}$ does not contain $\{0\}$ as an accumulation point \cite{KKFA08} so that spreading becomes just impossible  when the system size $n$ goes to infinity.

Similar behaviour  to those expected in the class of models considered here should also occur in other non random systems
for example quasiperiodic system \cite{AA81}. Our universal conclusion is that complete spreading of wave packet cannot occur when the linear spectrum of the system is purely discrete and when nonlinearities are taken into account. In some sense, Anderson localization is not completely suppressed by non linearities as believed in early papers.

This work should suggest new questions about the thermal behaviour of the same models at very low temperature when the average amplitude of the thermal fluctuations becomes smaller than the crossover amplitude where KAM tori appears. Then the system should exhibit only few slowly diffusive chaotic spots ( or chaotic breathers) surrounded by an ocean of regular non diffusive quasi periodic fluctuations. Such a situation should be associated by a dramatic drop of the thermal conductivity so that complete thermalization may  become impossible within reasonable times. Numerical experiments of fast quenching in some nonlinear systems   has revealed the spontaneous formation of time periodic)Discrete Breathers slowing down the thermalization though in these examples,  the linear spectrum was not discrete but completely absolutely continuous \cite{TA96}. The same kind of numerical experiments in systems with discrete spectrum (for example random) should dramatically enhance the effect. This may be also the situation in real glasses  where many questions remains unanswered  \cite{YC21} (despite that the linear phonon spectrum in such bond disorder system should not be purely discrete but with anabsolutely continuous part corresponding to low frequency and long wave length acoustic phonons).

I acknowledge  G. Tsironis  and  T. Bountis  for interesting discussions and  questions suggesting improvements for this manuscript.  I also thank again T. Bountis for inviting me to this meeting  and Ch. Skokos for communicating their very interesting recent papers.

\end{document}